\documentclass{aa}
\usepackage{graphicx}
\usepackage{natbib}
\usepackage{txfonts}
\begin{document}
\title{Planet formation in highly inclined binaries}

\subtitle{}

\titlerunning{Planets in inclined binaries}

\author{F. Marzari
        \inst{1},
        P. Th\'ebault
        \inst{2}
        and
        H. Scholl
        \inst{3}
        }

\offprints{F. Marzari}

\institute{
              Dipartimento di Fisica, University of Padova, Via Marzolo 8,
              35131 Padova, Italy\\
              \email{marzari@pd.infn.it}
         \and
             LESIA, Observatoire de Paris,
             Section de Meudon,
             F-92195 Meudon Principal Cedex, France\\
             \email{philippe.thebault@obspm.fr}
         \and
              Laboratoire Cassiop\'ee, Universit?? de Nice Sophia Antipolis, CNRS,
              Observatoire de la C\^ote d'Azur, B.P. 4229, F-06304 Nice Cedex, France\\
             \email{Hans.Scholl@oca.eu}
             }

\date{Received XXX ; accepted XXX}

\abstract
{}
{We explore planet formation in binary systems around the central star
where the protoplanetary
disk plane is highly inclined with respect to the companion star orbit. 
This might be the most frequent scenario for binary separations larger than 40 AU, 
according to Hale (1994). We focus on planetesimal accretion 
and compute average impact velocities in the habitable region and
up to 6 AU from the primary.
}
{Planetesimal trajectories are computed within the
frame of the restricted 3--body problem determined by the
central star, the companion star and massless planetesimals.
Relative velocities are computed and 
interpreted in terms of accreting or eroding impacts.
}
{We first show that, for binary inclinations higher than 10 degrees,
planetesimals evolve, at a first approximation, in a
gas-free environment.
Planetesimal accretion is confined around the central
star to a region determined by two main parameters, firstly by the
mutual inclination between the binary plane and the disk, and, secondly,
by the binary eccentricity.
}
{The onset of large mutual inclinations between planetesimals 
due to the nodal randomization causes an increase in the 
relative velocity. The chances for a successful planet 
accumulation process depend on the
balance between the timescale for node randomization and that
of planetesimal accretion. When the binary semimajor 
axis is larger than 70 AU, planet formation appears possible even for eccentric binaries 
(up to 0.4). For lower 
binary separations the region where planetesimals accumulate into protoplanets
shrinks consistently. When the mutual inclination between 
the binary plane and that of the planetesimal disk is larger
than $40^{\circ}$ the Kozai mechanism strongly inhibits 
planetesimal accumulation. 
}


\maketitle

\section{Introduction}

Planetary formation in binary systems is
a complex issue, since each step of the 
process can be affected in different ways by the companion perturbations.
Recent numerical studies \citep{mascho00,theb04,theb06,theb08,theb09,paard08,xie08}
have shown that one stage is particularly sensitive to the presence of the 
secondary star: the initial accretion of kilometer-sized planetesimals
(a review on this topic is given in \cite{nade}).
Indeed, the coupled effect of 
secular perturbations of the companion star and friction due to
gas in the nebulae induces a size-dependent phasing of orbits which
may lead to high impact velocities. This could slow down or even
halt the accretion process even in the terrestrial planet region
for a wide range of binary separations, i.e., 
up to $a_b\sim50\,$AU for high eccentricity systems
\citep[see for example Figs.8 and 9 in][]{theb06}.

However, these studies are based on the assumption that the planetesimal 
disk is coplanar to the stellar orbit. Even if this assumption might
intuitively seems reasonable, a systematic study by \cite{hale} 
on binary systems with solar-type components suggests that the spin of 
the two stars is aligned only for binary systems of 30--40 AU or less. 
Beyond this distance, the primary's equator, and thus a putative
planetesimal disk, appears to be randomly inclined
respect to the binary planet. 
As a consequence, the inclination between the binary's orbital plane
and the circumprimary disk is a parameter that has to be taken into
account when studying planetesimal accretion, at least
for systems with $a_b>30-40\,$AU.

We will focus in this paper on binaries with intermediate separations,
i.e. in the 40-100\,AU range, exploring the inclination as a
free parameter. Similarly to the studies for the coplanar case, the main
outcome we are interested on is the impact velocity distribution
within the planetesimal population, since this parameter
controls the fate of planetesimal collisions, either accretion or
erosion. For a significant mutual inclination between the 
binary orbital plane and the disk of planetesimals embedded in
the gaseous disk, 
the forced inclination
due to the companion star might be much more effective in increasing the 
relative velocities and halting planet formation. As the planetesimals 
decouple from the gaseous disk and evolve gravitationally, they would 
feel the binary perturbation and move into inclined orbits. 
The perturbations
of the companion star leads to a progressive randomization of
planetesimal node longitudes, starting from the 
outer region of the disk where the secular periods are shorter. 
The planetesimal disk gradually evolves into a cloud with an angular opening equal to 
twice the mutual initial inclination of the disk respect to the binary plane.
We explore in this paper the effects of the nodal randomization on the 
mutual relative velocities within the planetesimal 
swarm and 
on the accretion process. We also estimate the minimum inclination 
below which planet formation may occur in spite of
the binary inclination.

The paper is organized as follows: in Section 2. we 
show that the planetesimal dynamics perturbed 
by the companion star keeps the swarm  
out of the gaseous disk for most of the orbital period. 
This makes gas drag a negligible perturbation.
In Section 3. 
we describe the numerical model used to compute the 
planetesimals relative velocities.  Section 4. 
is devoted to the analysis of the impact 
velocities for different binary parameters.  
In Section 5. we derive limiting inclinations
for accretion at different binary separations. 
Finally, in Section 6. we summarize our results.

\section{Decoupling between gaseous disk and planetesimals}

\begin{figure}
\resizebox{\hsize}{!}{\includegraphics[angle=0]{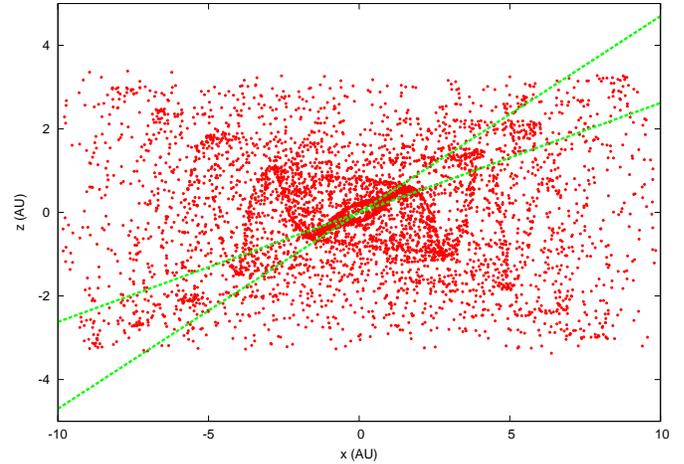}}
\caption[]{Planetesimal positions (red squares) in the x--z 
cartesian plane 
after $1 \times 10^5$ yrs from the beginning of their gravity dominated 
evolution. The gaseous disk (assuming a scale height  
$h = 0.05 (r/AU)^{(5/4)}$ AU)
is shown by green dashed lines. The orbital plane of the binary is 
in the x--y plane and the disk is inclined  of 
$i_m = 20^{\circ}$. 
}
\label{f1}
\end{figure}

Most recent studies of planetesimal accretion in a binary
environment \citep{mascho00,theb06,theb08,theb09,paard08}
have focused on the influence
of the gaseous component of the disk on particle dynamics.
However, the implicit assumption that the planetesimal swarm
is embedded in the gas disk is only valid if 
the disk is coplanar to the binary orbital plane. 
In this case, planetesimals feel a steady gas drag
and have their orbital evolution significantly affected
by frictional forces. 
However, if the companion star is on an inclined orbit with
respect to the disk mid--plane, the situation is dynamically
more complicated. 
Three possible scenarios can be envisaged
for the interactions between gas and planetesimals:

\begin{itemize}
\item Planetesimals form within the gas disk which remains 
a long--lived coherent entity in spite of the binary 
perturbations. Numerical simulations 
with constant viscosity and a polytropic equation of 
state performed 
by \cite{larw} with an SPH code suggest that a disk 
perturbed by an inclined companion star
maintains a coherent structure if the Mach number is 
lower than 30. It behaves like a rigid body preceding
at a rate $\omega_p$ given by:

\begin{equation}
\omega_p = - \frac{15 M_s R_d^3}{32 M_p D^3} cos (i_m) \Omega(R)
\end{equation}

where $M_s$ and $M_p$ are the masses of the secondary and primary star,
respectively, $D$ is the radius of the circular orbit of the 
binary, $i_m$ is the mutual inclination between the disk and 
the binary orbit, $\Omega(R)$ is the keplerian frequency
and  $R_d$ is the disk radius. 
This equation is derived under 
the simplified assumption that the 
disk has a constant density, but it is in general a good approximation
to more general cases. 
In this scenario, when the planetesimals reach the size 
(1--10 km in diameter) for which they evolve  
 under the dominating gravitational force of the two stars, 
they leave the disk plane because of the forced component in the  
inclination. Their orbits move in the binary orbital plane 
and their nodes circulate at different rates, depending on their semimajor
axis. Gas drag is probably not a significant perturbation 
in this scenario, since it affects planetesimal evolution only in the
fraction of time during which they cross the gaseous disk 
plane. This is clearly illustrated in
Fig.\ref{f1}, where we show the projection of the planetesimal
positions with respect to the gaseous disk when the inclination of 
the binary orbital plane is $i_m = 20^{\circ}$ with respect to the 
initial disk plane. Planetesimals spend most of their 
time out of disk where the gas density is negligible. 
According to our simulations, for $i_m = 30^{\circ}$  planetesimals
spend on average only 9\% of their time within one scale height
of the gaseous disk. This fraction increases to 13\% when
$i_m = 20^{\circ}$ and to 27\% when $i_m = 10^{\circ}$. 
As a consequence, we estimate that for $i_m \geq 20^{\circ}$
gas friction can be, to a first approximation, completely neglected
when computing planetesimals orbital evolution, while 
the $i_m = 10^{\circ}$ case appears as a limiting value
below which gas friction has to be taken into account.

\item The gaseous disk begins to warp and it is disrupted by the 
binary perturbations. It loses coherence and the gas
is dispersed in space.
According to \cite{larw} such
disruption by differential precession might affect extremely thin disks.
Also in this case, the planetesimals would evolve
in a gas-free environment. If the disk disrupts before the planetesimals detach from the
disk then this would be the most significant gas free case, where
planet formation would start from a disk of solid material made of
small planetesimal precursors which would evolve under gravity only.

\item As in the first case, the disk remains coherent
but it relaxes to the binary
plane on a timescale comparable to the viscous timescale 
\citep{larw}. If the process is fast due to a high 
viscosity of the disk, kilometer-sized planetesimals 
have not the time to form
before the disk relaxes to the binary plane. 
Planetesimals would then grow when their orbital plane, and 
that of the disk, are already 
aligned to that of the binary. In this case any information on the 
initial inclination would be lost and the system would evolve as 
a coplanar case \citep{mascho00,theb06,theb08,theb09,paard08}.

\end {itemize}

Apart from the case of fast relaxation, which possibly occurs in a minority
of cases with very high viscosity, 
in all other cases we expect
no or very weak coupling between
the gas disk and the planetesimals orbital evolution for
binaries with inclination $i\geq 10^{\circ}$.
Note that this low-$i$ case with gas drag has been
recently investigated by \citet{xie09}, who showed that
small inclinations between the binary and a circumprimary disc
might favor planetesimal accretion as compared to the fully
coplanar case.

For our numerical exploration, we will thus make the
simplifying assumption that planetesimals evolve in
a {\it gas--free} environment:
the gas--drag force, which introduces a de-phasing
of the planetesimal perihelia, does not come into play as in the 
2--D case introducing a de-phasing of the planetesimal 
perihelia. The evolution of the swarm can be described as a
pure gravitational N--body problem  \citep{theb06} where the 
relative impact velocity steadily increases because of both the 
de--phasing of perihelia and nodes. In the next section we will 
numerically compute the evolution of planetesimal relative 
velocities. 

\section{Numerical procedure}

Planetesimal trajectories are computed within the
frame of the restricted 3--body problem made of the
central star, the companion star and massless planetesimals.
We use the same code as in previous studies of the 2-D
case \citep[e.g.][]{theb06,theb08,theb09}, since this
code is 3-D in essence and can handle out of plane perturbers.
As already mentioned, the main parameter we are interested
in is the evolution of the average impact velocities
within the population of test planetesimals.
To that effect, our code has a build-in close encounter
search algorithm, which tracks down at each timestep all
2-body encounters, allowing to precisely compute the
relative velocity for each collision (a "collision" being
defined as a close encounter within an "inflated radius" 
equal to $3 \times 10^{-4}$ AU assigned
to each particle, see \citet{theb09} for more details).
The precision we obtain in our relative velocity 
estimate is of the order of 5m/s at 2AU.

These values of $\langle \Delta v \rangle$ have then
to be interpreted in terms of accreting or eroding impacts.
The limit between erosion and accretion is defined by
a threshold velocity $v*_{s1,s2}$, which depends on
the respective sizes $s_1$ and $s_2$ of the impactors, as well
as on the value of $Q*_{s1,s2}$, the threshold energy for
catastrophic fragmentation. Unfortunately, the parameter
$Q*$ is very poorly constrained and estimates found in
the literature can differ by up to more than 2 orders
of magnitude. We chose here a careful approach and consider
that $Q*$ is comprised between 2 limiting values for a "hard"
and "weak" prescription. This will in turn result in 2 
bracketing values for $v*_{s1,s2}$ \citep[see the discussion
in][for more details]{theb06}. 

The initial planetesimal swarm is made  of 15000 bodies 
initially set on a 
2--dimensional disk inclined by an angle 
$i_0$ with respect to the binary orbital plane. All the 
bodies in the disk are started on circular orbits 
with semimajor axis ranging from 0.8 to 6.5\,AU. 
All the nodal lines are parallel since all the bodies 
are clustered in a disk shape. 
The mass of the primary and secondary stars are fixed to
$1M_{\odot}$ and $0.5M_{\odot}$ respectively. The binary's
orbital parameters $a_b$, $e_b$ are chosen
as free parameters. $a_b$ ranges from 40 to 100 AU; beyond
those values the perturbations of the companion in the initial 
phases of planetesimal accretion are too weak. The binary
eccentricity $e_b$ assumes different values from 0 to 
0.4. The inclination $i_0$ varies from $0^{\circ}$ to 
$40^{\circ}$. For larger inclinations the Kozai mechanism
strongly inhibits planetesimal accumulation, as we will 
see in the following. 

 Our initial model setup is based on the assumption that 
initially the planetesimal swarm form a flat disk, in other
words that it is a dynamically "quiet" system, 
with all planetesimals $e_{free}$ and $i_{free}=0$.
For a gas rich environement this choice might be justified
by the fact that the progenitors of the km-sized planetesimals are coupled
to the gas and cannot have large $e_{free}$ \footnote{although this issue might be
more complex than this simple picture, depending on how planetesimals
are formed (see Discussion in Thebault et al.2006)}. In the present case,
this "decoupling" is more delicate to define, since it could either
be the consequence of the planetesimals vertical dispersion
around a coherent gas disc (case 1), or the consequence of
the gas disc dispersal (case 2). In each case, the relative timing between
the arrival of the "initial" kilometre-sized planetesimals
and the decoupling from the gas is difficult to pinpoint.
In a worst case scenario, we could have an initial orbital distribution
where some planetesimals have inclination close to $i_{forced}$ while others
are still around $i=0$. This would introduce a high initial
free relative velocity component that could not be erased with time
(contrary to the gas-rich case, see Fig.10 of Thebault et al.2006).
This difficult issue clearly exceeds the scope of this paper, but
our results should probably be taken as a lower estimate in terms
of inhibition of planetesimal accretion

\section{Results}

\subsection{Dynamical behaviour, encounter velocities}

\begin{figure}
\resizebox{\hsize}{!}{\includegraphics[angle=-90]{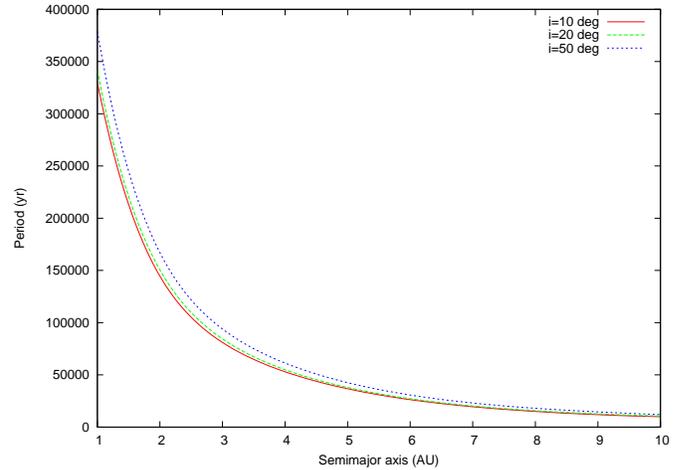}}
\caption[]{Circulation period of the nodel longitude
as a function of semimajor axis  
for planetesimals started on a disk around the 
primary star. The companion star has a semimajor axis
$a_b = 50$ AU, and eccentricity $e_b=0$ and different inclinations between the binary
orbital plane and the planetesimal disk are considered. The 
values of the circulation period are computed through direct
numerical integration of the equation of motion (3--body problem). 
}
\label{f2}
\end{figure}

\begin{figure}
\resizebox{\hsize}{!}{\includegraphics[angle=-90]{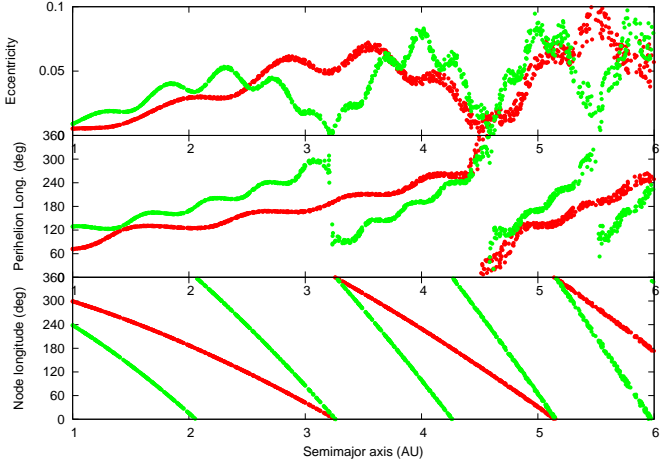}}
\caption[]{Distribution of the planetesimal orbital elements
at $t= 5 \times 10^4$ yr (green dots) and $t=10^5$ yr (red dots). 
The binary parameters are $i_b=30^o$, $e_b=0.2$ and $a_b=50$ AU. 
}
\label{forb}
\end{figure}

When the planetesimals feel the 
binary gravitational pull, their node longitude $\Omega$ 
starts precessing 
at a rate which strongly depends on the individual semimajor
axis of the bodies as shown in Fig.\ref{f2}. At the same time
the binary perturbations cause a growth of the eccentricity 
and a de--phasing of the perihelia. In Fig.\ref{forb} we 
illustrate the orbital distribution of the planetesimal swarm 
at $t = 5 \times 10^4$ and $t = 10^5$ yr when the 
binary orbital plane is inclined by $30^o$ respect to
the planetesimal disk and the eccentricity of the binary
orbit is 0.2 (the semimajor axis is 50 AU). 
The different timescales of perihelia and 
node circulation are manifest in the plot. The different 
degree of randomization of $\varpi$ and $\Omega$ makes 
a difficult task to predict the evolution of the 
relative impact velocity between the planetesimals.

\begin{figure}
\resizebox{\hsize}{!}{\includegraphics[angle=-90]{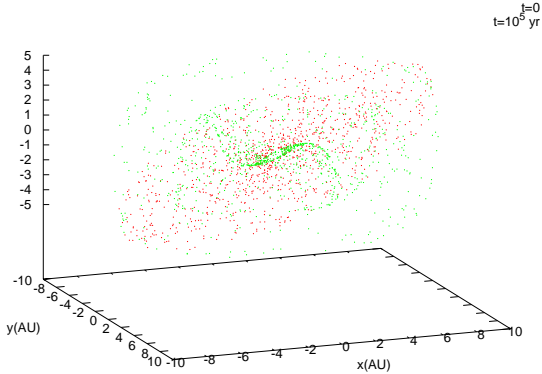}}
\caption[]{3--D spatial distributions of planetesimals at 
$t=0$ (red dots) and $t = 1 \times 10^5$ yr (green dots). The randomization of the 
nodal lines disrupt the coherence of the initial disk starting from 
the outer regions. The disk becomes an extended cloud. 
The binary orbital parameters are
$a_b = 50$ AU, $e_b=0.$ and $i_0 = 30^{\circ}$. 
}
\label{f3}
\end{figure}

The planetesimal disk moves to the binary orbital plane within 
one orbital period of the outer planetesimals and gradually 
looses coherence as a disk.  The nodes are randomized and 
the planetesimal Keplerian orbits
take them out of the disk plane.  
In Fig.\ref{f3} we illustrate 
this behaviour by plotting the positions of the 
planetesimals at $t=0$, when they are still grouped in a disk, 
and at $t = 1 \times 10^5$ yr when the randomization has 
disrupted their initial spatial configuration. We adopt this 
timescale since it is a higher limit for the initial planetesimal
accumulation process \citep[e.g.][]{liss93}

\begin{figure}
\resizebox{\hsize}{!}{\includegraphics[angle=0]{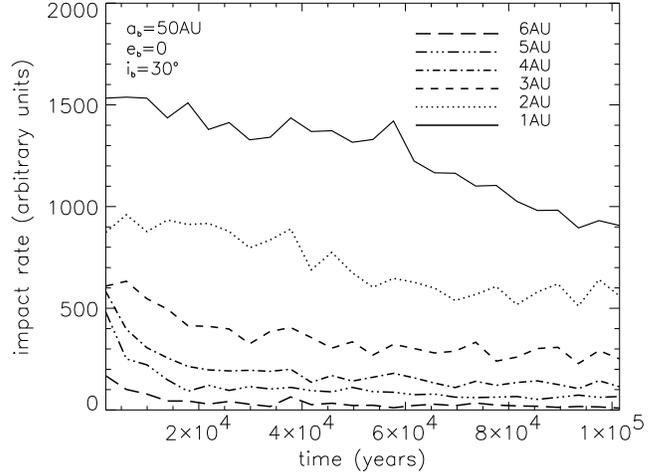}}
\caption[]{
Average impact rate in the planetesimal 
swarm as a function of time for the binary
configuration: $a_b=50$\,AU,$e_b=0$ and $i_b=30^{o}$.
Different curves refer to increasing radial distances from 
the primary star. 
The randomization of the 
node longitudes lead to a sparse planetesimal configuration
leading to a reduction of the encounter rate. 
}
\label{f4}
\end{figure}

The randomization of the node longitudes affects the dynamics of the
planetesimal population in two ways.
The first one is that the sparser distribution 
of the bodies in space leads to a lower impact rate in spite 
of the growth in the relative velocity. 
In Fig.\ref{f4} we plot the impact rate as a function of time 
and radial distance. It
shows a 
gradually declining trend as the nodes become more randomly distributed. 
This trend is more marked at larger distances from the primary 
star where the nodal randomization is faster. 
After $10^5$ years the impact rate is reduced by 55\% at 1 AU and it 
drops down by 94\% when $r = 6 AU$, independently 
of the planetesimal size. This percentage can be interpreted as
the fractional reduction of the impact rate compared to that of 
a planetesimal swarm around a single star. Around $t=0$ the impact
rate of our model is not yet affected by the binary perturbations
and it can be taken as representative of the impact rate around a single star.

The other, and more crucial effect is the progressive
increase of impact velocities, as can be clearly seen
in Fig.\ref{f5}. For values of $i_b\leq40^{o}$, this increase
is due to the combination of the particles small
free eccentricities $e_{fr}$
\footnote{The existence of a non-zero $e_{fr}$ component is
unavoidable. Indeed, for an unperturbed disc of
kilometer-sized planetesimals, equilibrium encounter velocities
are of the order of the bodies escape velocities, i.e. a few
m.s$^{-1}$, corresponding to $e_{fr}$ in the $10^{-5}-10^{-4}$ range}
and the large inclination oscillations
induced by the inclined companion star. Indeed, the small
random horizontal excursion due to $e_{fr}$ brings in contact
bodies having both $i$ and $\Omega$ values increasingly different
with time as  
the node
oscillations get tighter. This effect is of course more pronounced
in the outer regions of the disk, where the pull of the companion
star is felt more strongly. Note however that the steady $\langle \Delta v\rangle$
increase is observed everywhere in the 0.8-6\,AU region, it is just the
pace of this progressive increase which depends on radial distance.

For $i_b>40^{o}$ a fully different behaviour is observed and the 
Kozai oscillations come into play. The
eccentricity of the planetesimal orbits begin to grow while the
inclination decreases in order to keep the action:
\begin{equation}
\sqrt {h} = \sqrt{1 - e_{pl}^2} cos (I_{pl})
\label{eq:koza3}
\end{equation}
a constant of motion. The period of oscillation of $e_{pl}^2$ depends on
the semimajor axis of the planetesimal orbit and, as a consequence, large differences
build up in the distribution of orbital eccentricities of the swarm.
This leads to much higher impact velocities as is clearly shown in Fig.\ref{f6}.

\subsection{Effect on planetesimal accretion}

\begin{figure}
\resizebox{\hsize}{!}{\includegraphics[angle=0]{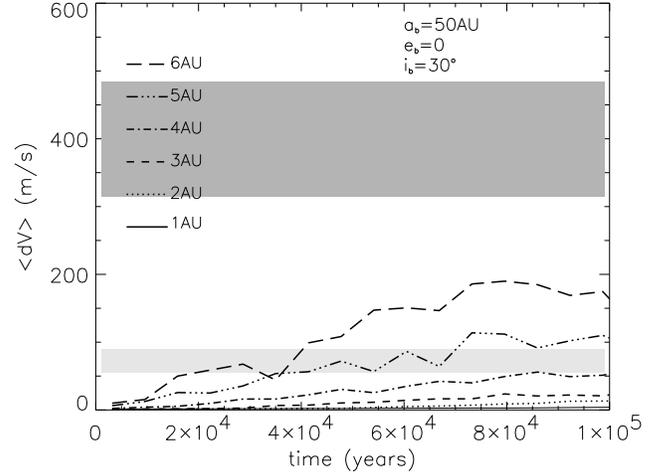}}
\caption[]{Average impact rate in the planetesimal 
swarm as a function of time at different radial distances 
from the primary star. The grey bands show the erosion limit
for planetesimals 1-10 km in size (lower band) and 10-50 km 
(upper band). The initial inclination between the planetesimal
disk and the binary orbit is $i_b = 30^{\circ}$.
}
\label{f5}
\end{figure}

As illustrated in Fig.\ref{f5} and discussed above, the increase
of impact velocities is an unavoidable consequence of 
the node randomization due to the companion star's perturbation
that affects all the simulated 0.8 to 6\,AU region.
However, in spite of this undesired effect, planetesimals might still 
undergo accretion and form planets. 
The critical condition is that the timescale for both the 
mutual velocity growth and impact rate reduction 
should be long compared to the accretion timescale.
More precisely, $\langle \Delta v \rangle$ 
have to stay at a low, accretion-friendly value long enough so that large 
planetesimals have enough time to build up. When the
high velocity regime is reached, the growing objects have reached
a $v*_{s1,s2}$ value high enough to have accreting impacts
despite higher $\langle \Delta v \rangle$. 
Of course, studying this effect in detail would require to
follow the evolution of the planetesimal size distribution in addition
to their dynamical one. This is unfortunately beyond the reach of
deterministic N-body codes \footnote{The size evolution of
a planetesimal population, under the effects of accreting, cratering and fragmenting
impacts, can only be treated in statistical particle-in-a-box codes for which the
dynamical modelling is necessarily very simplified}.
We shall thus adopt here a simplified criterion and consider the time at which
an averaged $\langle v* \rangle$ is reached for two cases:
a "small planetesimals" case
with $s_{min}=1$km and $s_{max}=10$km and a "large planetesimals"
case with $s_{min}=10$km and $s_{max}=50$km, assuming that planetesimal
sizes follow a centered Gaussian distribution between $s_{min}$ and
$s_{max}$. These two limiting $\langle v* \rangle$ values are indicated
by light and dark grey areas in Fig.\ref{f5}, the width of these areas
being due to our careful definition of $v*_{s1,s2}$ as being comprised
between two extreme values for hard and weak material (see Section 3).

As can be seen in Fig.\ref{f5}, for our example case
with $a_b=50$ AU, $e_b=0$ and $i_b = 30^{\circ}$, 
the whole system remains accretion-friendly for a population of large $\geq 10$km bodies
for the duration of the simulation, i.e. $10^{5}$ years, 
a conservative timescale for runaway growth.
For kilometer-sized planetesimals, the situation is less favorable and the $a\geq 5\,$AU
region becomes hostile to kilometre-sized planetesimal accretion after
$\sim 5\times 10^{4}$ years. In these regions, planet growth can proceed only
if in a few $10^{4}$ years planetesimals can grow large enough to
have accreting impacts in a $\langle \Delta v \rangle \sim 50-100$m.s$^{-1}$
environment. Note however that even $if$ planetesimal accretion is possible,
it can probably not lead to the same runaway growth as around an unperturbed 
single star \citep{korte}. 
Indeed, the increase of the impact velocity, even if it cannot stop
accretion, significantly slows it down by decreasing the value of
the gravitational focusing factor onto growing objects.
\citep[see the detailed discussion in][]{theb06}.
Past this initial planetesimal growth stage, it is difficult to predict the 
evolution of the swarm at farther stages when big planetesimals 
and planetary embryos will collide at these high relative velocities. Large relative inclinations
might be produced within the planetary system. \cite{quinta} have 
simulated the formation of terrestrial planets in $\alpha$ Centauri 
and found that planets may indeed be formed with large mutual inclinations if 
the orbital plane of the binary is inclined respect to that of 
the planetary embryos. However, their simulations start from a coherent  
and flat disk 
of protoplanets, while the growing protoplanets might have already
developed significant inclinations by the time they collide, according 
to our scenario. This suggests that the final phase of planet formation in 
inclined binaries may need additional investigation.  

\begin{figure}
\resizebox{\hsize}{!}{\includegraphics[angle=0]{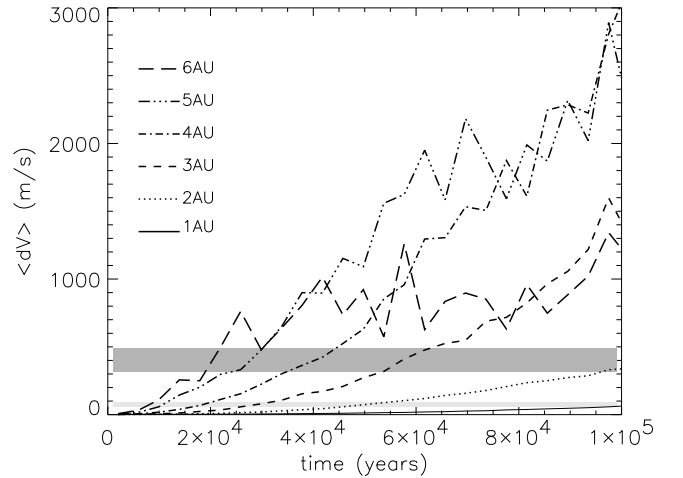}}
\caption[]{As in Fig.\ref{f5} for an initial inclination 
of the binary equal to $i_b=40^{\circ}$.
}
\label{f6}
\end{figure}

The situation is very different in the $i_b > 40^{\circ}$ cases where
Kozai oscillations dominate the planetesimal dynamics. 
As expected, these cases are much more hostile to planetesimal accretion.
As can be seen in Fig.\ref{f6} (for $a_b=50$ AU, $e_b=0$ and $i_b = 40^{\circ}$),
after $10^{5}$years the impact velocities reach values beyond the erosion 
limit for both "small" and "large" planetesimals in the whole
$a\geq 2$\,AU region.
Like in the non Kozai case, the rate at which $\langle \Delta v \rangle$
grows strongly depends on the radial distance. As an example, the $\geq 5\,$AU
region becomes hostile to the accretion of 1-10 km bodies after only
a few $10^{3}$years, whereas it takes almost $10^{5}$years for this 
to be true at $\sim 1\,$AU.

\cite{theb06} derived analytically the timescale before the onset of 
large impact velocities between planetesimals 
in planar eccentric binary systems as a function of the binary
parameters. In a gas free environment they computed the degree of 
perihelia randomization required to give high relative velocities 
and the time needed to reach it. 
Even if the inclined 
case may appear similar because accretion occurs in a gas free 
environment and the 
relative velocities grow because of the node randomization, it
is not possible to derive a similar analytical expression. 
It is a complex task 
to estimate how the relative encounter velocity depends on the degree of
node dispersion since this is a full 3--D problem. 
In addition, in this scenario both nodes and 
perihelia, once dispersed, contribute to the grow of the 
planetesimal relative velocities. 

\section{Role of the binary configuration}

In the previous section we have analyzed in details
the dynamical and accretional behaviour of 
a planetesimal population for two specific test binary configurations.
For pedagogical purposes, we chose cases with $e_b=0$ to focus on
the effect  of the binary inclination and, in particular, on the 
transition to a Kozai dominated regime when $i_b \geq 40^{\circ}$.
We now explore how these results depend on all the binary's orbital
parameters $a_b$, $e_b$ and $i_b$ (the mass ratio between the 2 stars
being kept constant and equal to 0.5).
For the sake of clarity, the accretion/erosion scenario for
each binary configuration will be summarized by a single simplified
parameter, the semimajor axis $a_l$ within which accretion is 
possible for the 1-10km planetesimal population at a threshold
timescale of $t_f \sim 5 \times 10^4 yr$ for the inner zone ranging
from $1-3$ AU and $t_f \sim 1 \times 10^5$ yr for the outer zone 
extending from $4-6$ AU.
We assume that planet formation is possible if 
$\langle \Delta v \rangle$ is lower than $v*_{s1,s2}$
for any $ t \leq \Delta t_f$. Under this condition
larger planetesimals can form and resist to higher 
velocity impacts. 
As an example, by inspecting Fig.\ref{f5} we can say that 
planetesimals beyond 5 AU reach the erosion regime before 
$1 \times 10^5$ yr while for 4 AU  $\langle \Delta v \rangle$ is still below the erosion
limit. The inner region within 3 AU has always impact velocities lower 
than $v*_{s1,s2}$ when $t < 5 \times 10^4$ yr. 

The choice of $ \Delta t_f $ is somewhat difficult and  arbitrary. 
Runaway growth in a planetesimal swarm around a single star
is supposed to start after about $10^4$ yrs while after 
$10^5$ yrs approximately 33\% of the disk mass is supposed to be
in runaway bodies, according to \cite{westu}. These values cannot be directly 
applied to the binary case since 1) the binary perturbations increase 
the relative velocities between the planetesimals accelerating the 
erosion rate 2) the collisional frequency decreases with time because of 
nodal dispersion. In this scenario it is difficult to derive
a reliable value for the $ \Delta t_f $ without knowing the details 
of the planetesimal accretion process. This would be possible only with 
a statistical code like the planet building code (\cite{wei}) which,
on the other hand, cannot  model the perturbations of 
a companion star.  Here we cautiously consider a value of 
$ \Delta t_f $ for the inner and outer region of the planetesimal 
disk which is larger than the runaway growth timescale and 
should somehow be a good estimate for the time required by planetesimals
to  
grow large enough to 
sustain further accretion into protoplanets.

\begin{figure}
\resizebox{\hsize}{!}{\includegraphics[angle=-90]{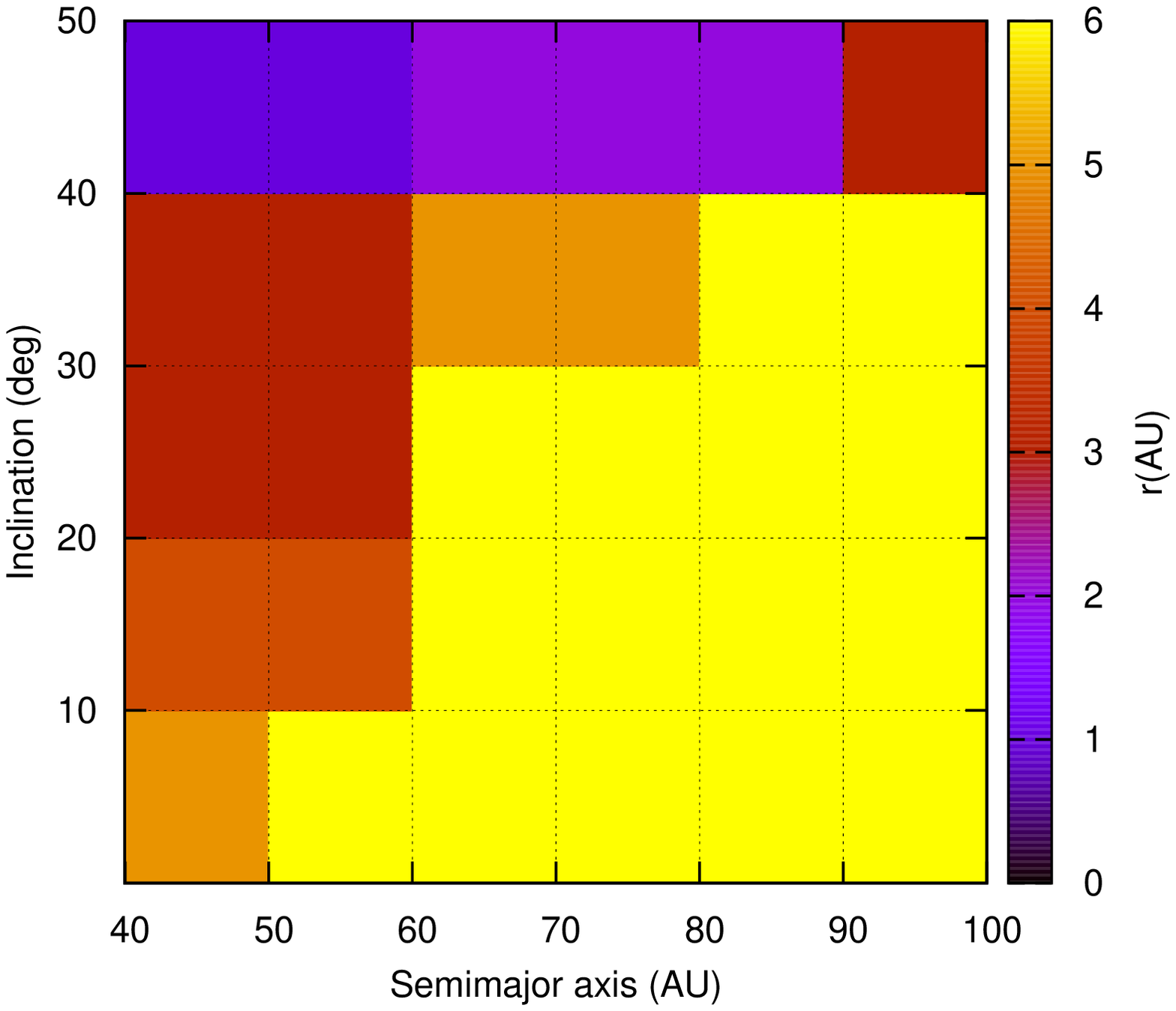}}
\resizebox{\hsize}{!}{\includegraphics[angle=-90]{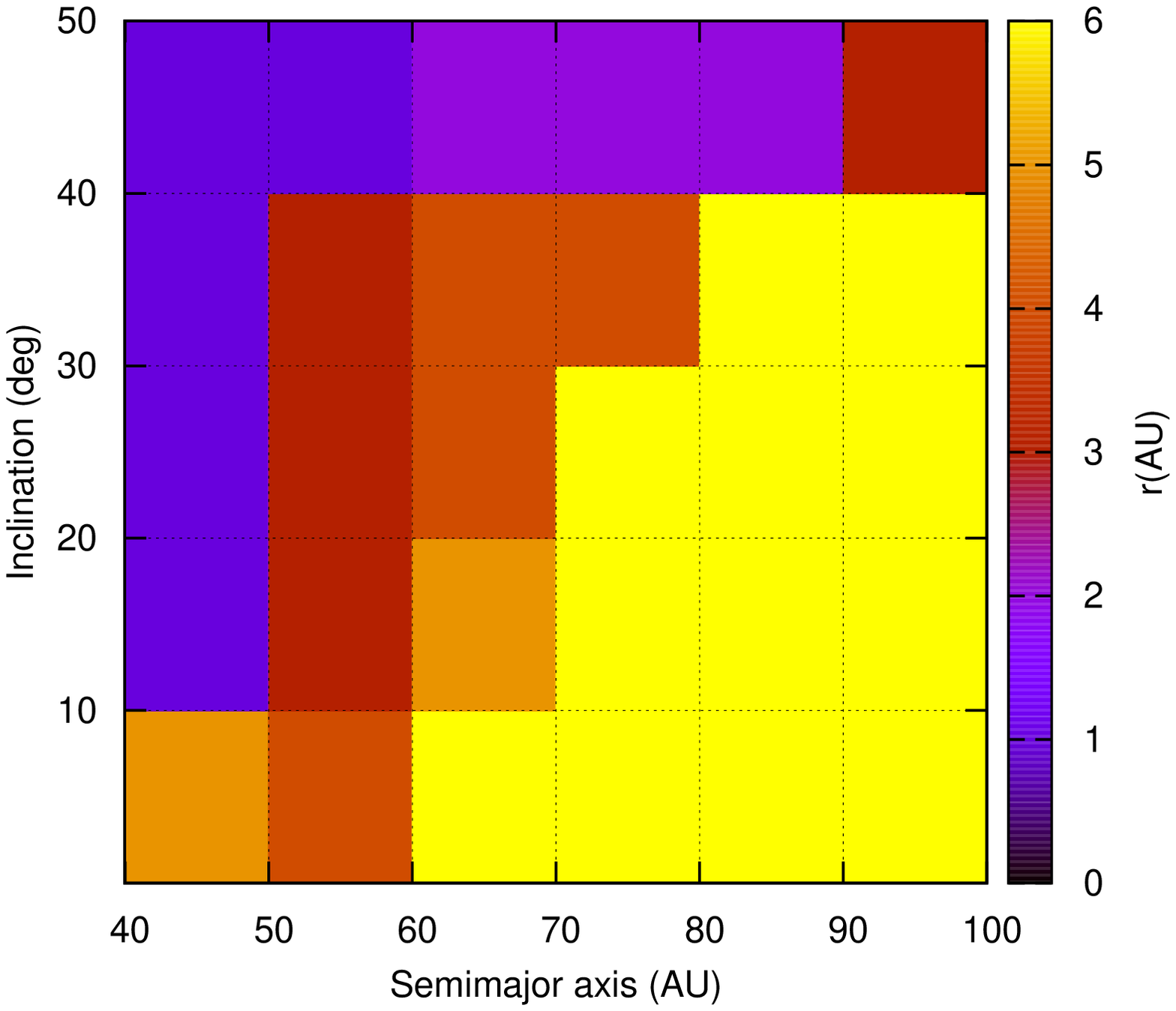}}
\resizebox{\hsize}{!}{\includegraphics[angle=-90]{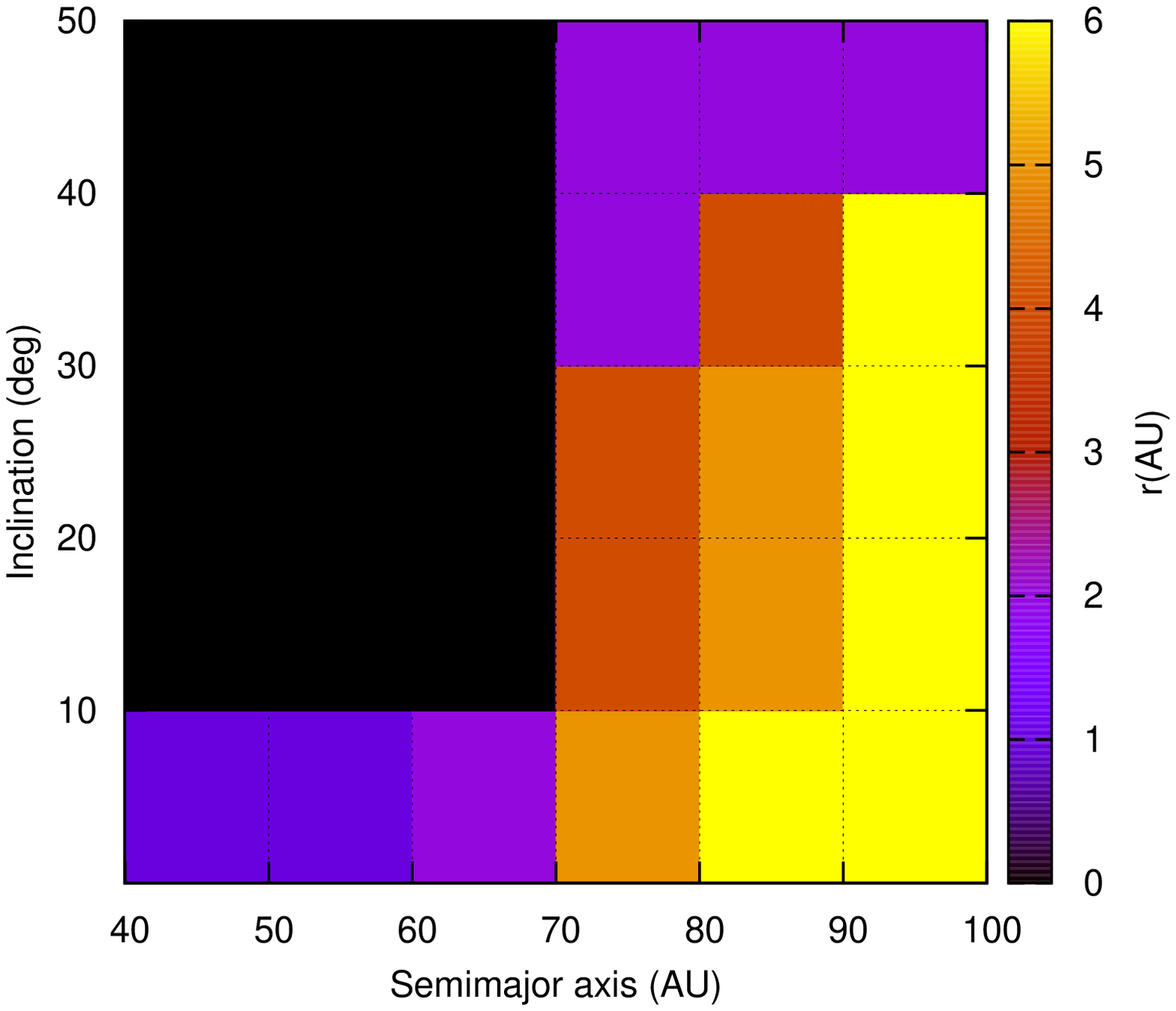}}
\caption[]{Maps showing the limiting values for accretion
$a_l$ as a function of ($a_b$, $i_b$). The top plot refers to the 
case with $e_b = 0.0$, the middle plot to $e_b = 0.2$, and
the lower plot to $e_b=0.4$. The color coding 
gives different values of $a_l$ (in AU), the limiting semimajor axis 
beyond which planetesimal accretion is possible. 
Each square of the map  refers to the lower value of 
the labels in the axes. The cases for $i_b = 0^o$ do not include
gas drag so they are only indicative.  
}
\label{f7}
\end{figure}

The outcome of this analysis is shown in the form of 2-dimensional 
maps. In Figs.\ref{f7} we plot $a_l$ vs. ($a_b$,$i_b$) for $e_b=0.0$, 
$e_b=0.2$ and 
$e_b=0.4$. The outcome for $i_b = 0^{\circ}$ is only a reference value 
since for low inclinations gas drag comes into play and it 
must be included in the numerical model \citep[for these low inclination
case, see the recent study by ][]{xie09}. 
It is noteworthy that in
the analysis of the data it never happens that accretion is possible 
beyond 4 AU and is prevented within 3 AU.  

When the companion star is on a circular orbit
(Fig.\ref{f7}$_a$), the  effect of inclination is noteworthy only for small values of $a_b$. The 
secular period of the nodes are short enough to perturb the disk on a 
timescale comparable to the accretion timescale only for $a_b \leq 50$AU. For
these small separations and $e_b=0$ planet formation is always possible
in the $\leq 3\,$AU region, except for the Kozai regime at $i_b \geq 40^{\circ}$ 
Beyond this point the swarm is quickly eroded because
of the enhanced relative velocities stirred up by the nodal randomization. 
At small value of  $a_b$ there is a strong dependence of $a_l$ on $i_b$ 
showing that at the origin of the shrinking of the planet formation 
zone there is the randomization of the nodes. 
For binary semimajor axes larger than 50 AU, the situation is much
more favorable to accretion, which can only be stopped, before $t_f$,  
in the Kozai regime with $i_b\geq 40^{\circ}$. 
This means that $a_b \sim 50$AU is approximately the border
value within which secular perturbations \emph{alone}
are efficient enough to affect planet accretion in the non Kozai regime.

For more eccentric binary orbits, the randomization of both
nodes and perihelia combine destructively and push the limit for
accretion at larger values of $a_b$. 
As an illustration, Fig.\ref{f7}$_c$ shows the situation for $e_b=0.4$.
In the black zone the relative velocity
is beyond the erosion value even at 0.8 AU from the primary star (the inner limit
for our planetesimal population). In 
these cases the formation of planets, in particular in
the outer regions of the disc, is 
strongly inhibited. For inclinations lower than $\sim 10^{\circ}$ some
accretion is possible within 1--2 AU. However, as already stated, our model for $i_b \leq 
10^{\circ}$ is less accurate since gas drag may in this case affect the evolution of
planetesimals. For $a_b$ larger than 70 AU we retrieve the dependence of 
$a_l$ on $i_b$ and at $a_b=90$ AU, planetesimal accumulation can only be perturbed
in the Kozai regime.

\section{Conclusions}

In this paper we explore the effect of high ($i_b \geq 10^{\circ}$) binary inclination on the 
planetesimal accretion process. The main outcomes of our work are:

\begin{itemize}

\item The gaseous disk and the planetesimals decouple because of 
the forced inclination of the companion star. As a consequence, 
planetesimal accumulation should occur in an almost gas free environment for 
inclinations approximately larger than $10^{\circ}$.

\item The progressive randomization of the planetesimal node
longitudes lead to the dispersion of the planetesimal disk 
that expand into a cloud of bodies surrounding the star. 
The sparser configuration leads to a significant reduction 
in the collisional rate. 

\item The onset of large mutual inclinations among planetesimals 
causes an increase of the relative impact velocity that may halt
the planet formation process. This effect is particularly 
strong for $i_b \geq 40^{\circ}$ where the Kozai mechanism 
comes into play. Below this value, planetesimal accretion
might be possible, preferentially in the regions closest
to the primary star, depending on the value of $i_b$ 

\item The possibility of planetesimal accumulation depends on the 
balance between the timescale of node randomization and that
of planetesimal accretion. For a binary on circular orbit, 
the value of $a_b$ around which this balance occurs is around 50 AU. Within this
value the secular perturbations are fast enough to induce 
large relative velocity on a timescale shorter 
than the typical timescale for planetesimal accumulation. Outside this 
limit planetesimals have probably enough time to growth big enough to
sustain high velocity impacts.  

\item When the binary is on an eccentric orbit, the randomization of 
nodes and periastra can lead to destructive collisions for binary 
separations up to 70 AU. 

\item The dispersion of planetesimals in the nodal longitude, in those
cases where the accretion is effective and lead to protoplanets, 
possibly leads to planetary systems a) that form on longer timescales 
because of the reduction of the accretion rate b) on highly mutually inclined 
orbits. 

\end{itemize}

For binary semimajor axes much larger than those we considered in this
paper the nodal longitude randomization becomes much longer.
As a consequence, the first stages of planetary formation will probably
proceed unaffected by the companion's perturbations: the planetesimal disk
is coherent during accumulation into rocky planets and core of
giant planets.
In this scenario, planets with 
a significant inclination respect to the binary orbit can form.
For intial inclinations larger than $\sim 40^o$ the Kozai cycles may lead,
on a long timescale,
a planet into a highly eccentric orbit which, at the same time, 
is very inclined with respect to the equator of the primary. 
This occurs because during the cycle the 
inclination respect to the binary plane is significantly decreased at the 
eccentricity peak leading the planet far from the equatorial plane
of the primary.  This might explain 
the observed orbit of HD 80606b (\cite{wum,gill,pont}). 

\begin{acknowledgements}
We thank the referee for his useful comments and suggestions that
helped to improve the paper. 
\end{acknowledgements}


\end{document}